\documentclass[aps,prl,10pt,twocolumn,groupedaddress, amsmath, showpacs,showkeys]{revtex4-1}
\usepackage{graphicx}
\usepackage{color}
\usepackage{bm}
\usepackage{dsfont}
\usepackage{braket}

\begin{document}


\title{Two-laser dynamic nuclear polarization with semiconductor electrons:\\
feedback, suppressed fluctuations, and bistability near two-photon resonance}


\author{A.~R.~Onur}
\author{C.~H.~van der Wal}
\affiliation{Zernike Institute for Advanced Materials, University of Groningen, 9747AG Groningen, The Netherlands}

\date{\today}

\begin{abstract}
We present how optical coherent population trapping (CPT) of the spin of localized semiconductor electrons stabilizes the surrounding nuclear spin bath via the hyperfine interaction, resulting in a state which is more ordered than the thermal equilibrium state. We find distinct control regimes for different signs of laser detuning and examine the transition from an unpolarized, narrowed state to a polarized state possessing a bistability. The narrowing of the state yields slower electron spin dephasing and self-improving CPT. Our analysis is relevant for a variety of solid state systems where hyperfine-induced dephasing is a limitation for using electron spin coherence.
\end{abstract}


\maketitle

A localized electron spin coupled to nuclear spins in a solid allows for studying the dynamics of mesoscopic spin ensembles. It forms a realization of the Gaudin (central spin) model \cite{gaudin1976} with the number of spins ranging from $\sim$10--$10^6$. From an application perspective the isolated dynamics of the electron spin is interesting as it can be used for quantum information processing. In thermal equilibrium the nuclear spins act as a source of dephasing for the electron spin. Optical orientation of the electron spin can be used to prepare out-of-equilibrium nuclear spin states via dynamic nuclear polarization (DNP) \cite{lampel1968,brown1996,urbaszek2013}. In turn, polarized nuclear spins induce an energy shift for the electron spin states, which can be described as an effective magnetic (Overhauser) field. DNP can also reduce thermal fluctuations in the nuclear spin polarization, which increases the electron spin dephasing time. This can be done either by creating a large nuclear spin polarization or by squeezing the polarization into a narrowed distribution \cite{coish2004}. Significant achievements have been made for both cases via electron transport, electron spin resonance, and optical preparation techniques \cite{urbaszek2013,coish2004,greilich2007,reilly2008,vink2009,xu2009,bluhm2010,ladd2010,togan2011,latta2011,urbaszek2013,hansom2014,stanley2014}. We present here how optical coherent population trapping (CPT) of localized semiconductor electrons stabilizes the surrounding nuclear spin bath in a state which is more ordered than the thermal equilibrium state.

CPT is the phenomenon where two-laser driving from the electron spin states to a common optically excited state displays --on exact two-photon resonance-- a suppression of optical excitation due to destructive quantum interference in the dynamics \cite{arimondo1976}, and is a key effect in quantum information processing \cite{fleischhauer2005}. Its sharp spectral feature allows for highly selective control over absorption and spontaneous emission of light. With atoms this has been applied in selective Doppler and sideband cooling \cite{tannoudji1988,morigi2000,roos2000}. Similarly, in semiconductors the CPT resonance can selectively address localized electrons that experience a particular Overhauser field \cite{stepanenko2006,issler2010,korenev2011}. This can lead to trapping of the combined electron-nuclear spin system in a dark state which was demonstrated as a measurement-based technique for reducing uncertainty of the nuclear spin state around a nitrogen vacancy center \cite{togan2011}.

The CPT-based control scheme we propose relies on an autonomous feedback loop, existing for detuned lasers only, and does not require measurement or adaptation of control lasers \cite{stepanenko2006}. Earlier work found such a feedback loop in an effective two-level description of a driven three-level $\Lambda$ system \cite{korenev2011}. We use a full description of the $\Lambda$ system dynamics and uncover distinct control regimes for different signs of the detuning and examine the transition from an unpolarized, narrowed state for blue-detuned lasers to a polarized state possessing a bistability for red-detuned lasers. With a stochastic approach that was previously used in the context of electron spin resonance experiments \cite{danon2008,vink2009} we analyze the evolution of thermalized nuclear spins to a state of reduced entropy. We also contrast our method with earlier work on quantum dots that relied on DNP from hyperfine interaction for the hole in the optically excited state \cite{xu2009,shi2013}. Our analysis assumes DNP that is driven by hyperfine contact interaction for the ground state electron \cite{abragam1961} and this gives different features, easily distinguishable in experiment. Our method thus expands the established CPT technique for coherent electron spin preparation and manipulation \cite{yale2013} to one that can also improve the electron spin dephasing time by nuclear spin preparation. The prerequisites are a high nuclear spin temperature and a non-zero electron spin temperature (ensuring bidirectional DNP). In example calculations we use parameters that approach (in order of magnitude) the values that apply to localized electrons in GaAs \cite{urbaszek2013}.


\begin{figure}
\centering
\includegraphics[width=\columnwidth]{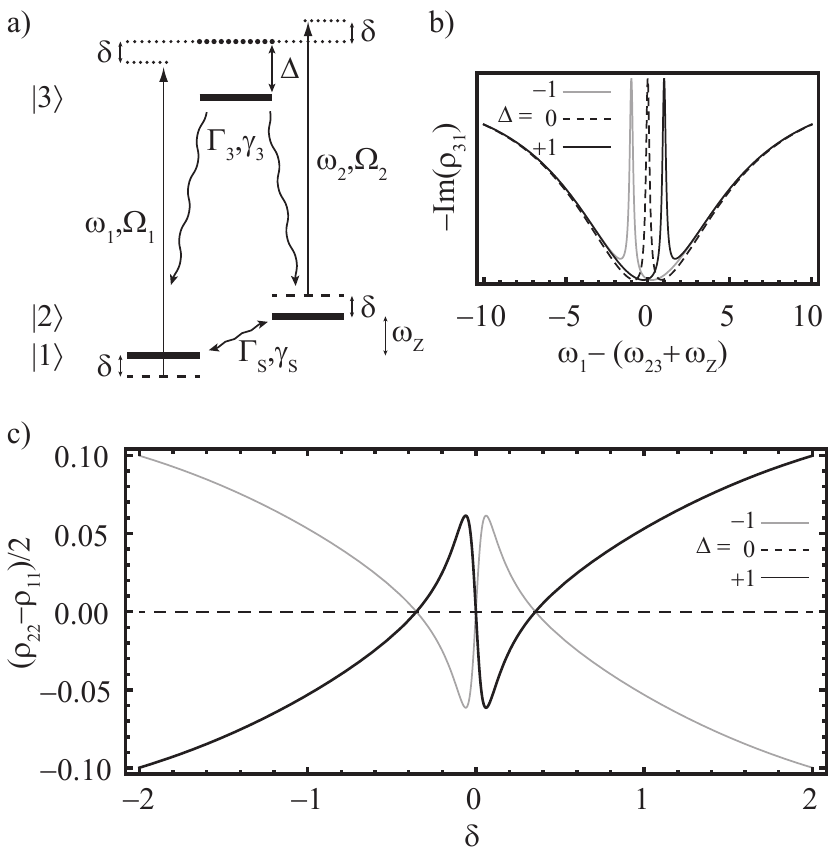}
\caption{(a) Schematic of energies and shifts of the electronic three-level system. Thick black lines are the (not Overhauser shifted) spin states $\ket{1}$, $\ket{2}$ and optically excited state $\ket{3}$. $\Gamma_{s}$, $\gamma_{s}$ and $\Gamma_{3}$ ,$\gamma_{3}$ are spin and excited state decay and decoherence rates, respectively. Two lasers (frequencies $\omega_{1}$ and $\omega_{2}$) couple to the system with Rabi frequencies $\Omega_{1}$ and $\Omega_{2}$, excited state detuning $\Delta$, and Overhauser shift $\delta$ (see further main text). (b) Conventional depiction of CPT (here for $\delta = 0$ Overhauser shift) showing the narrow CPT resonance within a broader absorption line. Laser 1 scans over the resonance while laser 2 is held fixed at $\omega_{2}=\omega_{23}+\Delta$, for detunings $\Delta$ as labeled. (c) Electron spin polarization as a function of Overhauser shift $\delta$, with lasers fixed at $\omega_{1}=\omega_{13}+\Delta$ and $\omega_{2}=\omega_{23}+\Delta$. In (b) and (c) results are presented as elements $\rho_{ij}$ of the steady-state density matrix. Parameters are normalized with respect to $\Gamma_{3}\equiv1$: $\gamma_{3}=10,\Gamma_{s}=10^{-4},\gamma_{s}=10^{-3},\Omega_{1}=\Omega_{2}=0.5$.}
\label{fig:lambda}
\end{figure}

Figure~\ref{fig:lambda}(a) presents the electronic part of our model: a $\Lambda$~system with spin states $\ket{1}$ and $\ket{2}$ that each have an optical transition to state $\ket{3}$. Nuclear spin polarization gives an Overhauser shift $-(+)\hbar \delta$ of the state $\ket{1}$ ($\ket{2}$), and we assume the Overhauser shift of $\ket{3}$ to be negligible. The values of energy differences $\hbar\omega_{13}$ and $\hbar\omega_{23}$, and Zeeman splitting $\hbar\omega_{z}$ between these states are defined for $\delta=0$. Two laser fields with frequencies $\omega_{1}$ and $\omega_{2}$ (and Rabi frequencies $\Omega_{1}$ and $\Omega_{2}$) selectively drive the two transitions.
The decay and decoherence rates of the system are the spin flip rate $\Gamma_{s}$, excited state decay rate $\Gamma_{3}$, spin decoherence rate $\gamma_{s}$ and excited state decoherence rate $\gamma_{3}$. We take all decay rates symmetric for the two electron spin states (for $\Gamma_{s}$ this implies temperature $k_B T >> \hbar \omega_z$), to avoid needless complication of the discussion, but our conclusions remain valid for the non-symmetric case. For modeling the CPT effects we directly follow Ref.~\cite{fleischhauer2005}. The Appendix specifies this in our notation. For this system, CPT occurs for driving at two-photon resonance (TPR, \textit{i.e.}, for $\delta=0$, $\omega_{1} = \omega_{2}+\omega_{z}$). In the conventional picture CPT is presented as a reduced absorption when $\omega_{1}$ is scanned across the resonance while $\omega_{2}$ is fixed near resonance at single-laser detuning $\Delta$. At the TPR point, the system gets trapped in a dark state that equals (for ideal spin coherence) $\ket{\Psi} \propto \Omega_{2}\ket{1} - \Omega_{1}\ket{2}$. Figure~\ref{fig:lambda}(b) presents this for different $\Delta$ in terms of the system's steady-state density-matrix element $\rho_{13}$.

\begin{figure}
\centering
\includegraphics[width=\columnwidth]{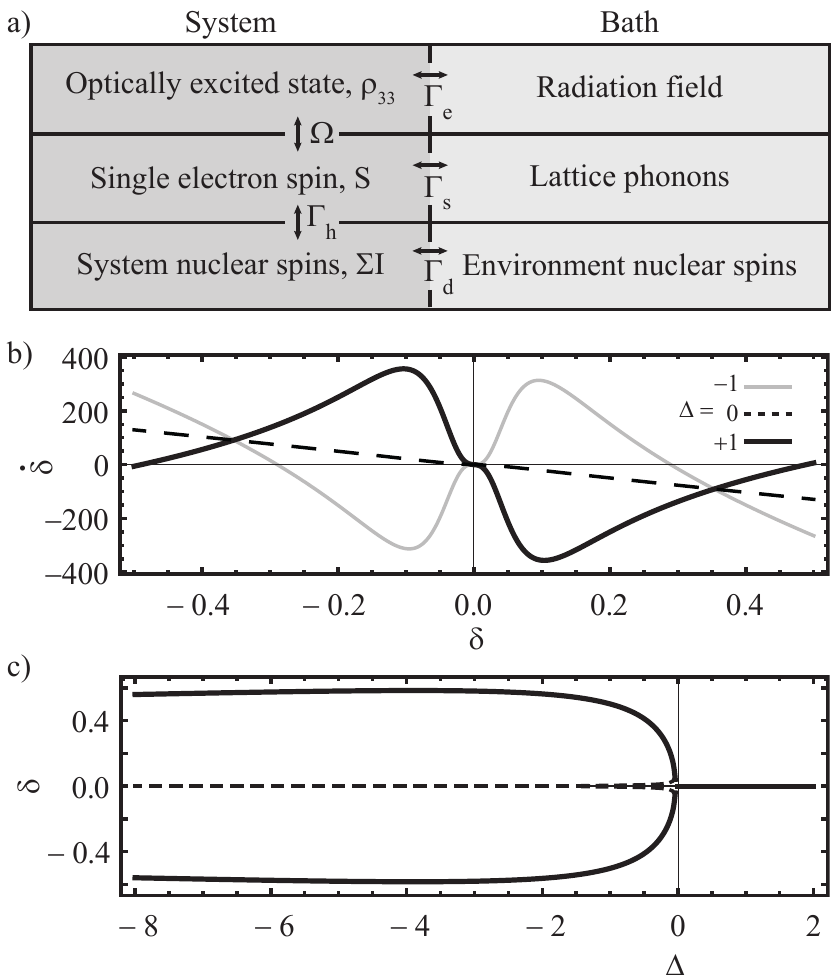}
\caption{(a) Overview of components and interactions of the laser-driven electron--nuclear-spin-ensemble system, with for each component its relaxation bath. Competition between the interactions and relaxation mechanisms govern the dynamics of the full system, see main text for details.
(b) The rate $\dot{\delta}$ as a function of the Overhauser shift $\delta$ (Eq.~(\ref{eq:feedbackdelta})) that is experienced by the electron near CPT conditions, for detunings $\Delta$ as labeled and $\overline{\Gamma}_{h}/\Gamma_{d}=0.01$.
(c) Thick solid (dashed) lines display the one or two (un)stable stationary $\delta$ values ($\dot{\delta}=0$) as a function of laser detuning $\Delta$.
The relaxation parameters and laser powers for (b) and (c) equal those of Fig.~\ref{fig:lambda}.}\label{fig:feedback}
\label{fig:opensystem}
\end{figure}

For our DNP analysis, however, we study CPT as a function of $\delta$ while the two lasers are tuned to exact TPR for $\delta = 0$. This is the electron's point of view on how a finite Overhauser shift breaks the ideal CPT condition, and the dependence on $\delta$ reflects the sharp spectral CPT feature. Figure~\ref{fig:lambda}(c) presents how this works out for the electron spin polarization, $(\rho_{22}-\rho_{11})/2$ in terms of the steady-state density matrix. The effect of a non-zero Overhauser shift is to break the TPR setting of the lasers. For $\Delta=0$ this has no effect on the spin population since $\delta$ drives both lasers away from resonance by an equal amount. For finite $\Delta$, however, the Overhauser shift leads to uneven detunings from the excited state, resulting in the electron spin population changing rapidly as a function of $\delta$ near TPR. Moreover, the electron spin population acquires a sign change as the sign of $\Delta$ is reversed. How this electron spin polarization as a function of $\delta$ drives DNP (which in turn will influence $\delta$) is the core of our further analysis. To this end, we consider the $\Lambda$-system to be embedded in the crystal lattice where it couples to nuclear spins within the electron wave function. We study the combined dynamics of the driven $\Lambda$~system and its surrounding nuclear spin bath, and also take into account that this nuclear spin bath in turn couples to other nuclear spins of the crystal that are not in contact with the electron, leading to leakage of nuclear spin polarization by spin diffusion (Fig.~\ref{fig:opensystem}(a)).


We first introduce relevant aspects of this hyperfine interaction. We concentrate on the common scenario where an external magnetic field is applied along $\hat{\mathbf{z}}$. This suppresses non-secular (not energy conserving) terms in the nuclear spin dipole-dipole interaction and we can approximate the nuclear spins to be frozen on the timescale of electron spin dynamics \cite{abragam1961, dyakonov1974, paget1977}. The hyperfine Hamiltonian has electron-nuclear flip-flop terms that describe the transfer of spin angular momentum along $\hat{\mathbf{z}}$ between the two systems (the Appendix provides a summary in our notation). For a single nuclear spin coupled to an electron, treated perturbatively, this results in the relaxation equation \cite{abragam1961}
\begin{equation}
\dot{\braket{I_{z}}}=-\Gamma_{h}\left(\braket{I_{z}}-\braket{\overline{I_{z}}}-
\frac{I^2+I}{S^2+S}\left[\braket{S_{z}}-\braket{\overline{S_{z}}}\right]\right)
\label{eq:basicdnp}.
\end{equation}
Here $I$ and $I_z$ are the nuclear spin quantum number and spin component along $\hat{\mathbf{z}}$, and similarly for electron spin $S$. The overbar indicates that the expectation value is taken at thermal equilibrium. The effective hyperfine relaxation rate $\Gamma_{h}$ is proportional to $\tau_{c}/(1+\omega_{z}^2\tau_{c}^2)$, which reflects how the electron spin correlation time $\tau_{c}$ determines the spectral density of the fluctuating hyperfine coupling \cite{abragam1961}. The quenching of optical excitation due to CPT near $\delta=0$ has an influence on $\Gamma_{h}$. In our model we take this into account by modulating the equilibrium hyperfine interaction rate $\overline{\Gamma}_{h}$ with the optical excitation rate obtained from the driven $\Lambda$-system dynamics of Fig.~\ref{fig:lambda} (see Appendix). Equation~(\ref{eq:basicdnp}) shows that $\braket{I_{z}}$ can be controlled by bringing the electron spin out of thermal equilibrium. By summing Eq.~(\ref{eq:basicdnp}) over all nuclei we can express the rate of change of $\delta$ as a function of $\delta$, forming a closed-loop system, which includes the dependence on the out-of-equilibrium electron spin polarization,
\begin{equation}
\dot{\delta}=-\Gamma_{h}\left[\delta-K\braket{S_{z}}\right]-\Gamma_{d}\delta,
\label{eq:feedbackdelta}
\end{equation}
where $K$ is a constant determined by the strength of the hyperfine coupling (see Appendix) and we used again the high temperature approximation $\braket{\overline{S_{z}}}=\braket{\overline{\delta}}=0$.
The last term of Eq.~(\ref{eq:feedbackdelta}) incorporates the loss of nuclear spin polarization by diffusion to the environment at a rate $\Gamma_d$ which we assume constant.

The polarization of the nuclear spin system is governed by the control dynamics of Eq.~(\ref{eq:feedbackdelta}). The dependence of this control on driving CPT for the electron is shown in Fig.~\ref{fig:opensystem}(b). Stable points are identified by $\dot{\delta}=0$ and $\frac{\partial\dot{\delta}}{\partial\delta}<0$. The dashed line represents the system driven by two lasers with $\Delta=0$, and has strong similarity with thermal equilibrium (no laser driving) since away from CPT there is no response of the electron spin polarization. The position of the stable point is at $\braket{\overline{\delta}}$, which we assumed zero. When the lasers are tuned to TPR for $\delta=0$ while having a finite detuning $\Delta$, two qualitatively different control regimes emerge. For the red-detuned case $\Delta=-1$ there are two stable points at $\delta \approx \pm 0.5$, and the nuclear spin system will thus display a bistability. For the blue-detuned case $\Delta=+1$, however, there is again one stable point at $\delta=0$. The transition between these two control regimes is shown in Fig.~\ref{fig:opensystem}(c) where the thick black lines represent the stable point(s) for a range of detunings $\Delta$. Even though the blue-detuned case displays the same stable point as the equilibrium case there is an enhanced response towards $\delta = 0$ for a region around this point. The effect of this gain becomes apparent when we study the stochastics of the nuclear spin polarization. Notably, the small plateaux in the traces of Fig.~\ref{fig:opensystem}(b) at $\delta = 0$ are due to the CPT suppression of $\Gamma_h$.

The stochastics of the nuclear spin polarization gives rise to the electron spin dephasing time that is observed in measurements, whether on an ensemble of $\Lambda$~systems \cite{sladkov2010} or by repeated measurements on a single system \cite{urbaszek2013}. In such cases each system experiences a different Overhauser shift, sampled from a probability distribution $P(\delta)$, and this directly translates into a distribution for the electron precession frequencies. This can be used to calculate the dephasing time $T_{2}^{*}$, indicating when information on the electron spin state has decayed to $1/\mathrm{e}$ of its initial value (see Appendix). The evolution of $P(\delta)$ under the control dynamics of Eq.~(\ref{eq:feedbackdelta}) can be described by a Fokker-Planck equation \cite{danon2008,danon2010}, in the continuum limit where the number of nuclear spins $N\gg1$,
\begin{equation}
\dot{P}=\frac{2}{N}\frac{\partial}{\partial\delta}\left(-\dot{\delta}P+\frac{\delta_{\text{max}}^{2}}{N}\frac{\partial}{\partial\delta}\left[\Gamma_{d}+\Gamma_{h}\right]P\right).
\label{eq:fokkerplanck}
\end{equation}
Here $N$ is the number of system nuclear spins and $\delta_{\text{max}}$ is the Overhauser shift for complete nuclear spin polarization (for simplicity, we describe the dynamics in the approximation where $N$ spins with $I$$=$$\frac12$ couple to the electron with equal strength \cite{urbaszek2013}). Without laser driving Eq.~(\ref{eq:feedbackdelta}) gives $\dot{\delta}=-(\Gamma_{d}+\Gamma_{h})\delta$ and the steady state solution to Eq.~(\ref{eq:fokkerplanck}) is a Gaussian with standard deviation $\sigma_{\delta}=\delta_{\text{max}}/\sqrt{N}$, as expected in thermal equilibrium. With laser driving the control gain becomes nonlinear, as in Fig. \ref{fig:opensystem}(b), and we evaluate the steady-state solution $P_{ss}(\delta)$ numerically (see Appendix).
With Eq.~(\ref{eq:fokkerplanck}) we can study the evolution of the initial thermalized distribution $\overline{P}(\delta)$ while laser control is imposed via Eq.~(\ref{eq:feedbackdelta}). The initial distribution depends on $N$ and $\delta_{\text{max}}$. For our example calculations we take $N=10^5$, $\delta_{\text{max}}=16.3$ and $K=10\delta_{max}/3$ (see Appendix), representing the donor-bound electron in GaAs \cite{sladkov2010} which has $\Gamma_{3}\approx1$~GHz.

\begin{figure}
\centering
\includegraphics[width=\columnwidth]{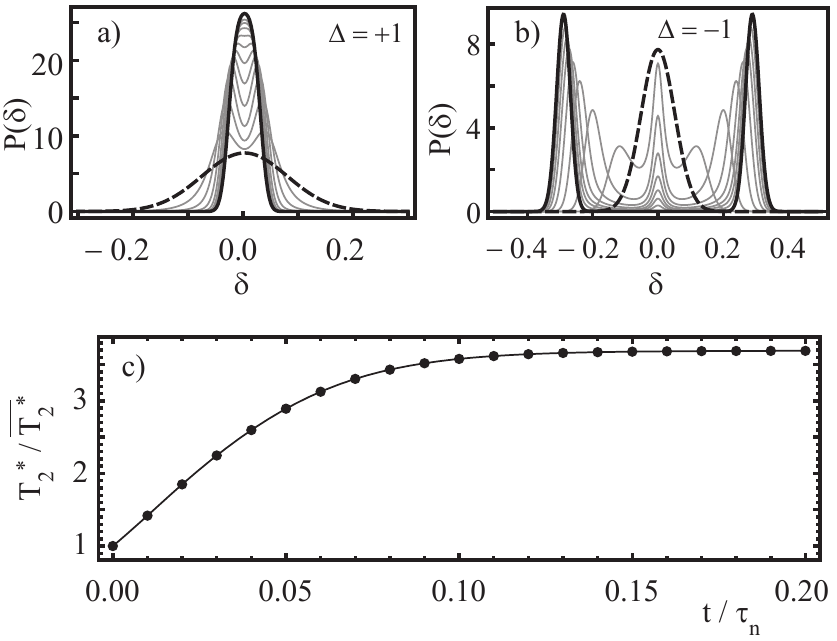}
\caption{Time evolution of $P(\delta)$ (probability distribution for $\delta$ values) for a nuclear spin bath with $N=10^5$, for cases that correspond to the curves in Fig.~\ref{fig:opensystem}(b), with $\Omega_{1}=\Omega_{2}=0.5$ and $\overline{\Gamma}_{h}/\Gamma_{d}=0.01$. In (a) and (b) the dashed lines show the same initial (Gaussian) distribution at thermal equilibrium (before laser driving is switched on), black lines show the final steady-state distribution. The sign of the detuning $\Delta$ determines whether the driven system has mono- or bistable behavior. Panel (c) shows the improvement in electron spin dephasing time corresponding to the sequence of curves in (a).}
\label{fig:evolution}
\end{figure}

The evolution of $P(\delta)$ corresponding to the response functions from Fig.~\ref{fig:feedback}(b) is depicted in Fig.~\ref{fig:evolution}(a,b). For the blue-detuned case $P(\delta)$ gets narrowed and focusses around the stable point $\delta=0$, while for the red-detuned case $P(\delta)$ splits apart and in the steady state it is divided between two stable points. During evolution the rate of change of $P(\delta)$ is at first lagging at $\delta=0$, causing the central dip in the gray lines of Fig. \ref{fig:evolution}(a) and the central peak in Fig. \ref{fig:evolution}(b). This is due to the suppressed hyperfine relaxation rate $\Gamma_h$ at CPT resonance. At long time scales this effect smoothes out.


A thermodynamic interpretation of this narrowing effect is that when the driven $\Lambda$ system is detuned from TPR, optical excitation converts low entropy laser light to higher entropy fluorescence light, resulting in an entropy flux away from the electron system. In turn, the electron acts as a controller on the nuclear spins, removing entropy from the spin bath and providing increased state information of the nuclear spins. Because the slow dynamics of the nuclei this effect is sustained after laser control is turned off, giving an enhanced dephasing time for subsequent electron spin manipulation. The evolution of $T_{2}^{*}$ calculated from $P(\delta)$ as in Fig.~\ref{fig:evolution}(a) is presented in Fig.~\ref{fig:evolution}(c), where the evolution time is expressed in units of the nuclear spin diffusion time $\tau_{n}=1/\Gamma_{d}$ (on the order of seconds to minutes). The nuclear spin bath attains a stable state with an increase in $T_{2}^{*}$ of a factor of $\approx3.7$ in $0.2\tau_{n}$. While this increase is moderate for the GaAs parameters used, it can be much more significant for systems with weaker nuclear spin diffusion (which can also be the case for GaAs when this is suppressed due to a Knight shift \cite{deng2005}). Notably, the resulting $P_{ss}(\delta)$ does not change with variation of $\overline{\Gamma}_{h}$ and $\Gamma_{d}$ provided their ratio remains fixed. For the system nuclear spins this represents the ratio of coupling strength to the controller (electron spin) and the environment (Fig.~\ref{fig:feedback}(a)).

\begin{figure}
\centering
\includegraphics[width=\columnwidth]{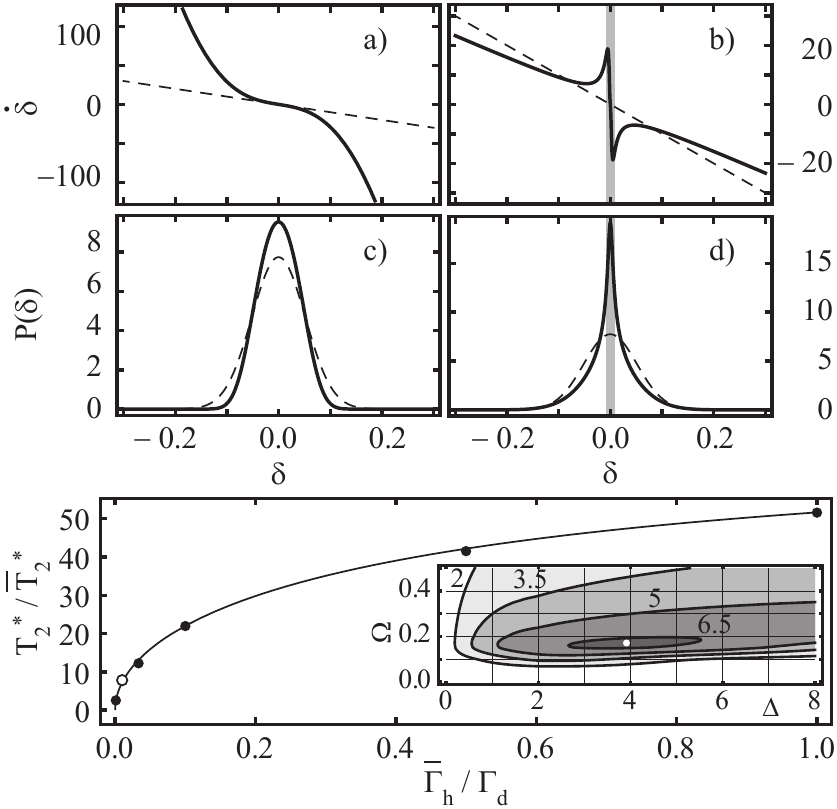}
\caption{Traces of driving rate $\dot{\delta}$ as a function of $\delta$ (black lines in panels a,b) and their respective effect on the nuclear spin distributions (c,d), for Rabi frequencies $\Omega=2$ (a,c) and $\Omega=0.1$ (b,d). In (a,b) the dashed line is $\Gamma_{d}\delta$, representing the nuclear spin flip rate due to spin diffusion. In (c,d) the dashed line is the same (note different scale) nuclear spin probability distribution function at thermal equilibrium for $N=10^5$ and $\overline{\Gamma}_{h}/\Gamma_{d}=0.01$. The black line is the steady-state distribution under laser driving at detuning $\Delta=+1$ and Rabi frequencies $\Omega=2$ (c) and $\Omega=0.1$ (d). The gray area in (b,d) highlights the narrowing range. For low laser powers the driving curve (b) shows a steep response at $\delta=0$ that acts as a strong force towards 0 for $\delta$ values around this point, and causes strong narrowing. The range over which narrowing takes place, however, is too small to cover the initial distribution. (e) Optimal $T_{2}^{*}$ improvement as a function of $\overline{\Gamma}_{h}/\Gamma_{d}$. The simulated values (dots) reveal the dependence $(\alpha/(1+\Gamma_{d}/\overline{\Gamma}_{h}))^{1/2}$, where $\alpha=73$ fits to this particular simulation (black line). Inset: Improvement factor in $T_{2}^{*}$ for a range of detunings and laser powers at $\overline{\Gamma}_{h}/\Gamma_{d}=0.01$. The white dot marks the optimum where $T_{2}^{*}/\overline{T}_{2}^{*} = 6.75$.}
\label{fig:powerdependence}
\end{figure}

Figure~\ref{fig:powerdependence} presents how the narrowing mechanism performs for different laser powers. At high power (Fig.~\ref{fig:powerdependence}(a,c), $\Omega_{1}=\Omega_{2}\equiv\Omega=2$) the power broadening of the CPT resonance quenches the hyperfine rate $\Gamma_h$ over a wide range around $\delta=0$. This results in a weak response and the narrowing is only effective at the tails of the initial $P(\delta)$. At lower power (Fig.~\ref{fig:powerdependence}(b,d), $\Omega=0.1$) there is a strong response around $\delta=0$, indicating strong narrowing. However this does not extend far enough to include the tails of the initial $P(\delta)$. The $T_{2}^{*}$ improvement factor in both cases is minor, only 1.38 and 1.63 respectively. Optima are found at moderate laser powers. Figure~\ref{fig:powerdependence}(e) depicts the optimum values as a function of $\overline{\Gamma}_{h}/\Gamma_{d}$ where dots are calculated values. The inset shows how such an optimum is found from a map of $T_{2}^{*}/\overline{T}_{2}^{*}$ for a range of laser powers and detunings for $\overline{\Gamma}_{h}/\Gamma_{d}=0.01$ (open circle in main figure). We find a square root dependence for the optima, i.e. $T_{2}^{*}/\overline{T}_{2}^{*}\propto (1/(1+\Gamma_{d}/\overline{\Gamma}_{h}))^{1/2}$. This reflects that at the optimum the response can be approximated as linear ($\dot{\delta} \propto \delta$) over the width of the final distribution $P_{ss}(\delta)$, which is then approximately Gaussian.


In Ref.~\cite{xu2009} a similar narrowing effect has been described and demonstrated for a quantum dot. The authors attribute it to the non-collinear hyperfine coupling for the hole spin in the optically excited state, while our result is based on electron-nuclear spin coupling. For paramagnetic defects, in general, either type of hyperfine coupling may dominate. To distinguish the two in experiment we point out two characteristics that are different and readily measurable. Firstly, the transition from narrowing to a regime of bistability with changing sign of the detuning only occurs for our model. Secondly, the narrowing in Ref.~\cite{xu2009} improves with increasing power while for our model there is a particular laser power that gives the optimal narrowing (Fig.~\ref{fig:powerdependence}(e)).

In conclusion we have presented a method that integrates CPT control of an electron spin with stabilizing control over the nuclear spin polarization around the electron. The time evolution of the open system contains an autonomous feedback loop, which stabilizes the nuclear spin bath in a more ordered configuration without requiring adaptation of the control fields. The effects we have discussed are readily measurable since the transmission of the laser beams tuned central on a narrow CPT line increases when the electron spin dephasing time increases. Hence, the narrowing of the nuclear spin polarization distribution directly translates to enhanced laser transmission over time (or equivalently, in a reduced signal when detecting fluorescence). Our method should be applicable to a wide range of spin defects in solid state.

\begin{acknowledgments}
We thank D.~O'Shea, J.~P.~de~Jong, J.~Sloot and, A.~U.~Chaubal for valuable discussions, and acknowledge financial support from FOM, NWO and an ERC Starting Grant.
\end{acknowledgments}

\bibliography{onur_main}

\begin{thebibliography}{32}%
\makeatletter
\providecommand \@ifxundefined [1]{%
 \@ifx{#1\undefined}
}%
\providecommand \@ifnum [1]{%
 \ifnum #1\expandafter \@firstoftwo
 \else \expandafter \@secondoftwo
 \fi
}%
\providecommand \@ifx [1]{%
 \ifx #1\expandafter \@firstoftwo
 \else \expandafter \@secondoftwo
 \fi
}%
\providecommand \natexlab [1]{#1}%
\providecommand \enquote  [1]{``#1''}%
\providecommand \bibnamefont  [1]{#1}%
\providecommand \bibfnamefont [1]{#1}%
\providecommand \citenamefont [1]{#1}%
\providecommand \href@noop [0]{\@secondoftwo}%
\providecommand \href [0]{\begingroup \@sanitize@url \@href}%
\providecommand \@href[1]{\@@startlink{#1}\@@href}%
\providecommand \@@href[1]{\endgroup#1\@@endlink}%
\providecommand \@sanitize@url [0]{\catcode `\\12\catcode `\$12\catcode
  `\&12\catcode `\#12\catcode `\^12\catcode `\_12\catcode `\%12\relax}%
\providecommand \@@startlink[1]{}%
\providecommand \@@endlink[0]{}%
\providecommand \url  [0]{\begingroup\@sanitize@url \@url }%
\providecommand \@url [1]{\endgroup\@href {#1}{\urlprefix }}%
\providecommand \urlprefix  [0]{URL }%
\providecommand \Eprint [0]{\href }%
\providecommand \doibase [0]{http://dx.doi.org/}%
\providecommand \selectlanguage [0]{\@gobble}%
\providecommand \bibinfo  [0]{\@secondoftwo}%
\providecommand \bibfield  [0]{\@secondoftwo}%
\providecommand \translation [1]{[#1]}%
\providecommand \BibitemOpen [0]{}%
\providecommand \bibitemStop [0]{}%
\providecommand \bibitemNoStop [0]{.\EOS\space}%
\providecommand \EOS [0]{\spacefactor3000\relax}%
\providecommand \BibitemShut  [1]{\csname bibitem#1\endcsname}%
\let\auto@bib@innerbib\@empty
\bibitem [{\citenamefont {Gaudin}(1976)}]{gaudin1976}%
  \BibitemOpen
  \bibfield  {author} {\bibinfo {author} {\bibfnamefont {M.}~\bibnamefont
  {Gaudin}},\ }\href@noop {} {\bibfield  {journal} {\bibinfo  {journal}
  {Journal de Physique}\ }\textbf {\bibinfo {volume} {37}},\ \bibinfo {pages}
  {1087} (\bibinfo {year} {1976})}\BibitemShut {NoStop}%
\bibitem [{\citenamefont {Lampel}(1968)}]{lampel1968}%
  \BibitemOpen
  \bibfield  {author} {\bibinfo {author} {\bibfnamefont {G.}~\bibnamefont
  {Lampel}},\ }\href {\doibase 10.1103/PhysRevLett.20.491} {\bibfield
  {journal} {\bibinfo  {journal} {Phys. Rev. Lett.}\ }\textbf {\bibinfo
  {volume} {20}},\ \bibinfo {pages} {491} (\bibinfo {year} {1968})}\BibitemShut
  {NoStop}%
\bibitem [{\citenamefont {Brown}\ \emph {et~al.}(1996)\citenamefont {Brown},
  \citenamefont {Kennedy}, \citenamefont {Gammon},\ and\ \citenamefont
  {Snow}}]{brown1996}%
  \BibitemOpen
  \bibfield  {author} {\bibinfo {author} {\bibfnamefont {S.~W.}\ \bibnamefont
  {Brown}}, \bibinfo {author} {\bibfnamefont {T.~A.}\ \bibnamefont {Kennedy}},
  \bibinfo {author} {\bibfnamefont {D.}~\bibnamefont {Gammon}}, \ and\ \bibinfo
  {author} {\bibfnamefont {E.~S.}\ \bibnamefont {Snow}},\ }\href {\doibase
  10.1103/PhysRevB.54.R17339} {\bibfield  {journal} {\bibinfo  {journal} {Phys.
  Rev. B}\ }\textbf {\bibinfo {volume} {54}},\ \bibinfo {pages} {R17339}
  (\bibinfo {year} {1996})}\BibitemShut {NoStop}%
\bibitem [{\citenamefont {Urbaszek}\ \emph {et~al.}(2013)\citenamefont
  {Urbaszek}, \citenamefont {Marie}, \citenamefont {Amand}, \citenamefont
  {Krebs}, \citenamefont {Voisin}, \citenamefont {Maletinsky}, \citenamefont
  {H\"ogele},\ and\ \citenamefont {Imamoglu}}]{urbaszek2013}%
  \BibitemOpen
  \bibfield  {author} {\bibinfo {author} {\bibfnamefont {B.}~\bibnamefont
  {Urbaszek}}, \bibinfo {author} {\bibfnamefont {X.}~\bibnamefont {Marie}},
  \bibinfo {author} {\bibfnamefont {T.}~\bibnamefont {Amand}}, \bibinfo
  {author} {\bibfnamefont {O.}~\bibnamefont {Krebs}}, \bibinfo {author}
  {\bibfnamefont {P.}~\bibnamefont {Voisin}}, \bibinfo {author} {\bibfnamefont
  {P.}~\bibnamefont {Maletinsky}}, \bibinfo {author} {\bibfnamefont
  {A.}~\bibnamefont {H\"ogele}}, \ and\ \bibinfo {author} {\bibfnamefont
  {A.}~\bibnamefont {Imamoglu}},\ }\href {\doibase 10.1103/RevModPhys.85.79}
  {\bibfield  {journal} {\bibinfo  {journal} {Rev. Mod. Phys.}\ }\textbf
  {\bibinfo {volume} {85}},\ \bibinfo {pages} {79} (\bibinfo {year}
  {2013})}\BibitemShut {NoStop}%
\bibitem [{\citenamefont {Coish}\ and\ \citenamefont {Loss}(2004)}]{coish2004}%
  \BibitemOpen
  \bibfield  {author} {\bibinfo {author} {\bibfnamefont {W.~A.}\ \bibnamefont
  {Coish}}\ and\ \bibinfo {author} {\bibfnamefont {D.}~\bibnamefont {Loss}},\
  }\href {\doibase 10.1103/PhysRevB.70.195340} {\bibfield  {journal} {\bibinfo
  {journal} {Phys. Rev. B}\ }\textbf {\bibinfo {volume} {70}},\ \bibinfo
  {pages} {195340} (\bibinfo {year} {2004})}\BibitemShut {NoStop}%
\bibitem [{\citenamefont {Greilich}\ \emph {et~al.}(2007)\citenamefont
  {Greilich}, \citenamefont {Shabaev}, \citenamefont {Yakovlev}, \citenamefont
  {Efros}, \citenamefont {Yugova}, \citenamefont {Reuter}, \citenamefont
  {Wieck},\ and\ \citenamefont {Bayer}}]{greilich2007}%
  \BibitemOpen
  \bibfield  {author} {\bibinfo {author} {\bibfnamefont {A.}~\bibnamefont
  {Greilich}}, \bibinfo {author} {\bibfnamefont {A.}~\bibnamefont {Shabaev}},
  \bibinfo {author} {\bibfnamefont {D.~R.}\ \bibnamefont {Yakovlev}}, \bibinfo
  {author} {\bibfnamefont {A.~L.}\ \bibnamefont {Efros}}, \bibinfo {author}
  {\bibfnamefont {I.~A.}\ \bibnamefont {Yugova}}, \bibinfo {author}
  {\bibfnamefont {D.}~\bibnamefont {Reuter}}, \bibinfo {author} {\bibfnamefont
  {A.~D.}\ \bibnamefont {Wieck}}, \ and\ \bibinfo {author} {\bibfnamefont
  {M.}~\bibnamefont {Bayer}},\ }\href {\doibase 10.1126/science.1146850}
  {\bibfield  {journal} {\bibinfo  {journal} {Science}\ }\textbf {\bibinfo
  {volume} {317}},\ \bibinfo {pages} {1896} (\bibinfo {year}
  {2007})}\BibitemShut {NoStop}%
\bibitem [{\citenamefont {Reilly}\ \emph {et~al.}(2008)\citenamefont {Reilly},
  \citenamefont {Taylor}, \citenamefont {Petta}, \citenamefont {Marcus},
  \citenamefont {Hanson},\ and\ \citenamefont {Gossard}}]{reilly2008}%
  \BibitemOpen
  \bibfield  {author} {\bibinfo {author} {\bibfnamefont {D.~J.}\ \bibnamefont
  {Reilly}}, \bibinfo {author} {\bibfnamefont {J.~M.}\ \bibnamefont {Taylor}},
  \bibinfo {author} {\bibfnamefont {J.~R.}\ \bibnamefont {Petta}}, \bibinfo
  {author} {\bibfnamefont {C.~M.}\ \bibnamefont {Marcus}}, \bibinfo {author}
  {\bibfnamefont {M.~P.}\ \bibnamefont {Hanson}}, \ and\ \bibinfo {author}
  {\bibfnamefont {A.~C.}\ \bibnamefont {Gossard}},\ }\href {\doibase
  10.1126/science.1159221} {\bibfield  {journal} {\bibinfo  {journal}
  {Science}\ }\textbf {\bibinfo {volume} {321}},\ \bibinfo {pages} {817}
  (\bibinfo {year} {2008})}\BibitemShut {NoStop}%
\bibitem [{\citenamefont {Vink}\ \emph {et~al.}(2009)\citenamefont {Vink},
  \citenamefont {Nowack}, \citenamefont {Koppens}, \citenamefont {Danon},
  \citenamefont {Nazarov},\ and\ \citenamefont {Vandersypen}}]{vink2009}%
  \BibitemOpen
  \bibfield  {author} {\bibinfo {author} {\bibfnamefont {I.~T.}\ \bibnamefont
  {Vink}}, \bibinfo {author} {\bibfnamefont {K.~C.}\ \bibnamefont {Nowack}},
  \bibinfo {author} {\bibfnamefont {F.~H.~L.}\ \bibnamefont {Koppens}},
  \bibinfo {author} {\bibfnamefont {J.}~\bibnamefont {Danon}}, \bibinfo
  {author} {\bibfnamefont {Y.~V.}\ \bibnamefont {Nazarov}}, \ and\ \bibinfo
  {author} {\bibfnamefont {L.~M.~K.}\ \bibnamefont {Vandersypen}},\ }\href
  {\doibase 10.1038/nphys1366} {\bibfield  {journal} {\bibinfo  {journal}
  {Nature Physics}\ }\textbf {\bibinfo {volume} {5}},\ \bibinfo {pages} {764}
  (\bibinfo {year} {2009})}\BibitemShut {NoStop}%
\bibitem [{\citenamefont {Xu}\ \emph {et~al.}(2009)\citenamefont {Xu},
  \citenamefont {Yao}, \citenamefont {Sun}, \citenamefont {Steel},
  \citenamefont {Bracker}, \citenamefont {Gammon},\ and\ \citenamefont
  {Sham}}]{xu2009}%
  \BibitemOpen
  \bibfield  {author} {\bibinfo {author} {\bibfnamefont {X.}~\bibnamefont
  {Xu}}, \bibinfo {author} {\bibfnamefont {W.}~\bibnamefont {Yao}}, \bibinfo
  {author} {\bibfnamefont {B.}~\bibnamefont {Sun}}, \bibinfo {author}
  {\bibfnamefont {D.~G.}\ \bibnamefont {Steel}}, \bibinfo {author}
  {\bibfnamefont {A.~S.}\ \bibnamefont {Bracker}}, \bibinfo {author}
  {\bibfnamefont {D.}~\bibnamefont {Gammon}}, \ and\ \bibinfo {author}
  {\bibfnamefont {L.~J.}\ \bibnamefont {Sham}},\ }\href {\doibase
  10.1038/nature08120} {\bibfield  {journal} {\bibinfo  {journal} {Nature}\
  }\textbf {\bibinfo {volume} {459}},\ \bibinfo {pages} {1105} (\bibinfo {year}
  {2009})}\BibitemShut {NoStop}%
\bibitem [{\citenamefont {Bluhm}\ \emph {et~al.}(2010)\citenamefont {Bluhm},
  \citenamefont {Foletti}, \citenamefont {Mahalu}, \citenamefont {Umansky},\
  and\ \citenamefont {Yacoby}}]{bluhm2010}%
  \BibitemOpen
  \bibfield  {author} {\bibinfo {author} {\bibfnamefont {H.}~\bibnamefont
  {Bluhm}}, \bibinfo {author} {\bibfnamefont {S.}~\bibnamefont {Foletti}},
  \bibinfo {author} {\bibfnamefont {D.}~\bibnamefont {Mahalu}}, \bibinfo
  {author} {\bibfnamefont {V.}~\bibnamefont {Umansky}}, \ and\ \bibinfo
  {author} {\bibfnamefont {A.}~\bibnamefont {Yacoby}},\ }\href {\doibase
  10.1103/PhysRevLett.105.216803} {\bibfield  {journal} {\bibinfo  {journal}
  {Phys. Rev. Lett.}\ }\textbf {\bibinfo {volume} {105}},\ \bibinfo {pages}
  {216803} (\bibinfo {year} {2010})}\BibitemShut {NoStop}%
\bibitem [{\citenamefont {Ladd}\ \emph {et~al.}(2010)\citenamefont {Ladd},
  \citenamefont {Press}, \citenamefont {De~Greve}, \citenamefont {McMahon},
  \citenamefont {Friess}, \citenamefont {Schneider}, \citenamefont {Kamp},
  \citenamefont {H\"ofling}, \citenamefont {Forchel},\ and\ \citenamefont
  {Yamamoto}}]{ladd2010}%
  \BibitemOpen
  \bibfield  {author} {\bibinfo {author} {\bibfnamefont {T.~D.}\ \bibnamefont
  {Ladd}}, \bibinfo {author} {\bibfnamefont {D.}~\bibnamefont {Press}},
  \bibinfo {author} {\bibfnamefont {K.}~\bibnamefont {De~Greve}}, \bibinfo
  {author} {\bibfnamefont {P.~L.}\ \bibnamefont {McMahon}}, \bibinfo {author}
  {\bibfnamefont {B.}~\bibnamefont {Friess}}, \bibinfo {author} {\bibfnamefont
  {C.}~\bibnamefont {Schneider}}, \bibinfo {author} {\bibfnamefont
  {M.}~\bibnamefont {Kamp}}, \bibinfo {author} {\bibfnamefont {S.}~\bibnamefont
  {H\"ofling}}, \bibinfo {author} {\bibfnamefont {A.}~\bibnamefont {Forchel}},
  \ and\ \bibinfo {author} {\bibfnamefont {Y.}~\bibnamefont {Yamamoto}},\
  }\href {\doibase 10.1103/PhysRevLett.105.107401} {\bibfield  {journal}
  {\bibinfo  {journal} {Phys. Rev. Lett.}\ }\textbf {\bibinfo {volume} {105}},\
  \bibinfo {pages} {107401} (\bibinfo {year} {2010})}\BibitemShut {NoStop}%
\bibitem [{\citenamefont {Togan}\ \emph {et~al.}(2011)\citenamefont {Togan},
  \citenamefont {Chu}, \citenamefont {Imamoglu},\ and\ \citenamefont
  {Lukin}}]{togan2011}%
  \BibitemOpen
  \bibfield  {author} {\bibinfo {author} {\bibfnamefont {E.}~\bibnamefont
  {Togan}}, \bibinfo {author} {\bibfnamefont {Y.}~\bibnamefont {Chu}}, \bibinfo
  {author} {\bibfnamefont {A.}~\bibnamefont {Imamoglu}}, \ and\ \bibinfo
  {author} {\bibfnamefont {M.~D.}\ \bibnamefont {Lukin}},\ }\href {\doibase
  10.1038/nature10528} {\bibfield  {journal} {\bibinfo  {journal} {Nature}\
  }\textbf {\bibinfo {volume} {478}},\ \bibinfo {pages} {497} (\bibinfo {year}
  {2011})}\BibitemShut {NoStop}%
\bibitem [{\citenamefont {Latta}\ \emph {et~al.}(2011)\citenamefont {Latta},
  \citenamefont {Srivastava},\ and\ \citenamefont {Imamo\ifmmode~\breve{g}\else
  \u{g}\fi{}lu}}]{latta2011}%
  \BibitemOpen
  \bibfield  {author} {\bibinfo {author} {\bibfnamefont {C.}~\bibnamefont
  {Latta}}, \bibinfo {author} {\bibfnamefont {A.}~\bibnamefont {Srivastava}}, \
  and\ \bibinfo {author} {\bibfnamefont {A.}~\bibnamefont
  {Imamo\ifmmode~\breve{g}\else \u{g}\fi{}lu}},\ }\href {\doibase
  10.1103/PhysRevLett.107.167401} {\bibfield  {journal} {\bibinfo  {journal}
  {Phys. Rev. Lett.}\ }\textbf {\bibinfo {volume} {107}},\ \bibinfo {pages}
  {167401} (\bibinfo {year} {2011})}\BibitemShut {NoStop}%
\bibitem [{\citenamefont {Hansom}\ \emph {et~al.}(2014)\citenamefont {Hansom},
  \citenamefont {Schulte}, \citenamefont {Le~Gall}, \citenamefont {Matthiesen},
  \citenamefont {Clarke}, \citenamefont {Hugues}, \citenamefont {Taylor},\ and\
  \citenamefont {Atat{\"u}re}}]{hansom2014}%
  \BibitemOpen
  \bibfield  {author} {\bibinfo {author} {\bibfnamefont {J.}~\bibnamefont
  {Hansom}}, \bibinfo {author} {\bibfnamefont {C.~H.~H.}\ \bibnamefont
  {Schulte}}, \bibinfo {author} {\bibfnamefont {C.}~\bibnamefont {Le~Gall}},
  \bibinfo {author} {\bibfnamefont {C.}~\bibnamefont {Matthiesen}}, \bibinfo
  {author} {\bibfnamefont {E.}~\bibnamefont {Clarke}}, \bibinfo {author}
  {\bibfnamefont {M.}~\bibnamefont {Hugues}}, \bibinfo {author} {\bibfnamefont
  {J.~M.}\ \bibnamefont {Taylor}}, \ and\ \bibinfo {author} {\bibfnamefont
  {M.}~\bibnamefont {Atat{\"u}re}},\ }\href {\doibase doi:10.1038/nphys3077}
  {\bibfield  {journal} {\bibinfo  {journal} {Nature Physics}\ }\textbf
  {\bibinfo {volume} {advance online publication}} (\bibinfo {year} {2014}),\
  doi:10.1038/nphys3077}\BibitemShut {NoStop}%
\bibitem [{\citenamefont {{Stanley}}\ \emph {et~al.}(2014)\citenamefont
  {{Stanley}}, \citenamefont {{Matthiesen}}, \citenamefont {{Hansom}},
  \citenamefont {{Le Gall}}, \citenamefont {{Schulte}}, \citenamefont
  {{Clarke}},\ and\ \citenamefont {{Atat{\"u}re}}}]{stanley2014}%
  \BibitemOpen
  \bibfield  {author} {\bibinfo {author} {\bibfnamefont {M.~J.}\ \bibnamefont
  {{Stanley}}}, \bibinfo {author} {\bibfnamefont {C.}~\bibnamefont
  {{Matthiesen}}}, \bibinfo {author} {\bibfnamefont {J.}~\bibnamefont
  {{Hansom}}}, \bibinfo {author} {\bibfnamefont {C.}~\bibnamefont {{Le Gall}}},
  \bibinfo {author} {\bibfnamefont {C.~H.~H.}\ \bibnamefont {{Schulte}}},
  \bibinfo {author} {\bibfnamefont {E.}~\bibnamefont {{Clarke}}}, \ and\
  \bibinfo {author} {\bibfnamefont {M.}~\bibnamefont {{Atat{\"u}re}}},\
  }\href@noop {} {\bibfield  {journal} {\bibinfo  {journal} {ArXiv e-prints}\ }
  (\bibinfo {year} {2014})},\ \Eprint {http://arxiv.org/abs/1408.6437}
  {arXiv:1408.6437 [cond-mat.mes-hall]} \BibitemShut {NoStop}%
\bibitem [{\citenamefont {Arimondo}\ and\ \citenamefont
  {Orriols}(1976)}]{arimondo1976}%
  \BibitemOpen
  \bibfield  {author} {\bibinfo {author} {\bibfnamefont {E.}~\bibnamefont
  {Arimondo}}\ and\ \bibinfo {author} {\bibfnamefont {G.}~\bibnamefont
  {Orriols}},\ }\href@noop {} {\bibfield  {journal} {\bibinfo  {journal} {Lett.
  Nuovo Cimento}\ }\textbf {\bibinfo {volume} {17}},\ \bibinfo {pages} {333}
  (\bibinfo {year} {1976})}\BibitemShut {NoStop}%
\bibitem [{\citenamefont {Fleischhauer}\ \emph {et~al.}(2005)\citenamefont
  {Fleischhauer}, \citenamefont {Imamoglu},\ and\ \citenamefont
  {Marangos}}]{fleischhauer2005}%
  \BibitemOpen
  \bibfield  {author} {\bibinfo {author} {\bibfnamefont {M.}~\bibnamefont
  {Fleischhauer}}, \bibinfo {author} {\bibfnamefont {A.}~\bibnamefont
  {Imamoglu}}, \ and\ \bibinfo {author} {\bibfnamefont {J.~P.}\ \bibnamefont
  {Marangos}},\ }\href {\doibase 10.1103/RevModPhys.77.633} {\bibfield
  {journal} {\bibinfo  {journal} {Rev. Mod. Phys.}\ }\textbf {\bibinfo {volume}
  {77}},\ \bibinfo {pages} {633} (\bibinfo {year} {2005})}\BibitemShut
  {NoStop}%
\bibitem [{\citenamefont {Aspect}\ \emph {et~al.}(1988)\citenamefont {Aspect},
  \citenamefont {Arimondo}, \citenamefont {Kaiser}, \citenamefont
  {Vansteenkiste},\ and\ \citenamefont {Cohen-Tannoudji}}]{tannoudji1988}%
  \BibitemOpen
  \bibfield  {author} {\bibinfo {author} {\bibfnamefont {A.}~\bibnamefont
  {Aspect}}, \bibinfo {author} {\bibfnamefont {E.}~\bibnamefont {Arimondo}},
  \bibinfo {author} {\bibfnamefont {R.}~\bibnamefont {Kaiser}}, \bibinfo
  {author} {\bibfnamefont {N.}~\bibnamefont {Vansteenkiste}}, \ and\ \bibinfo
  {author} {\bibfnamefont {C.}~\bibnamefont {Cohen-Tannoudji}},\ }\href
  {\doibase 10.1103/PhysRevLett.61.826} {\bibfield  {journal} {\bibinfo
  {journal} {Phys. Rev. Lett.}\ }\textbf {\bibinfo {volume} {61}},\ \bibinfo
  {pages} {826} (\bibinfo {year} {1988})}\BibitemShut {NoStop}%
\bibitem [{\citenamefont {Morigi}\ \emph {et~al.}(2000)\citenamefont {Morigi},
  \citenamefont {Eschner},\ and\ \citenamefont {Keitel}}]{morigi2000}%
  \BibitemOpen
  \bibfield  {author} {\bibinfo {author} {\bibfnamefont {G.}~\bibnamefont
  {Morigi}}, \bibinfo {author} {\bibfnamefont {J.}~\bibnamefont {Eschner}}, \
  and\ \bibinfo {author} {\bibfnamefont {C.~H.}\ \bibnamefont {Keitel}},\
  }\href {\doibase 10.1103/PhysRevLett.85.4458} {\bibfield  {journal} {\bibinfo
   {journal} {Phys. Rev. Lett.}\ }\textbf {\bibinfo {volume} {85}},\ \bibinfo
  {pages} {4458} (\bibinfo {year} {2000})}\BibitemShut {NoStop}%
\bibitem [{\citenamefont {Roos}\ \emph {et~al.}(2000)\citenamefont {Roos},
  \citenamefont {Leibfried}, \citenamefont {Mundt}, \citenamefont
  {Schmidt-Kaler}, \citenamefont {Eschner},\ and\ \citenamefont
  {Blatt}}]{roos2000}%
  \BibitemOpen
  \bibfield  {author} {\bibinfo {author} {\bibfnamefont {C.~F.}\ \bibnamefont
  {Roos}}, \bibinfo {author} {\bibfnamefont {D.}~\bibnamefont {Leibfried}},
  \bibinfo {author} {\bibfnamefont {A.}~\bibnamefont {Mundt}}, \bibinfo
  {author} {\bibfnamefont {F.}~\bibnamefont {Schmidt-Kaler}}, \bibinfo {author}
  {\bibfnamefont {J.}~\bibnamefont {Eschner}}, \ and\ \bibinfo {author}
  {\bibfnamefont {R.}~\bibnamefont {Blatt}},\ }\href {\doibase
  10.1103/PhysRevLett.85.5547} {\bibfield  {journal} {\bibinfo  {journal}
  {Phys. Rev. Lett.}\ }\textbf {\bibinfo {volume} {85}},\ \bibinfo {pages}
  {5547} (\bibinfo {year} {2000})}\BibitemShut {NoStop}%
\bibitem [{\citenamefont {Stepanenko}\ \emph {et~al.}(2006)\citenamefont
  {Stepanenko}, \citenamefont {Burkard}, \citenamefont {Giedke},\ and\
  \citenamefont {Imamoglu}}]{stepanenko2006}%
  \BibitemOpen
  \bibfield  {author} {\bibinfo {author} {\bibfnamefont {D.}~\bibnamefont
  {Stepanenko}}, \bibinfo {author} {\bibfnamefont {G.}~\bibnamefont {Burkard}},
  \bibinfo {author} {\bibfnamefont {G.}~\bibnamefont {Giedke}}, \ and\ \bibinfo
  {author} {\bibfnamefont {A.}~\bibnamefont {Imamoglu}},\ }\href {\doibase
  10.1103/PhysRevLett.96.136401} {\bibfield  {journal} {\bibinfo  {journal}
  {Phys. Rev. Lett.}\ }\textbf {\bibinfo {volume} {96}},\ \bibinfo {pages}
  {136401} (\bibinfo {year} {2006})}\BibitemShut {NoStop}%
\bibitem [{\citenamefont {Issler}\ \emph {et~al.}(2010)\citenamefont {Issler},
  \citenamefont {Kessler}, \citenamefont {Giedke}, \citenamefont {Yelin},
  \citenamefont {Cirac}, \citenamefont {Lukin},\ and\ \citenamefont
  {Imamoglu}}]{issler2010}%
  \BibitemOpen
  \bibfield  {author} {\bibinfo {author} {\bibfnamefont {M.}~\bibnamefont
  {Issler}}, \bibinfo {author} {\bibfnamefont {E.~M.}\ \bibnamefont {Kessler}},
  \bibinfo {author} {\bibfnamefont {G.}~\bibnamefont {Giedke}}, \bibinfo
  {author} {\bibfnamefont {S.}~\bibnamefont {Yelin}}, \bibinfo {author}
  {\bibfnamefont {I.}~\bibnamefont {Cirac}}, \bibinfo {author} {\bibfnamefont
  {M.~D.}\ \bibnamefont {Lukin}}, \ and\ \bibinfo {author} {\bibfnamefont
  {A.}~\bibnamefont {Imamoglu}},\ }\href {\doibase
  10.1103/PhysRevLett.105.267202} {\bibfield  {journal} {\bibinfo  {journal}
  {Phys. Rev. Lett.}\ }\textbf {\bibinfo {volume} {105}},\ \bibinfo {pages}
  {267202} (\bibinfo {year} {2010})}\BibitemShut {NoStop}%
\bibitem [{\citenamefont {Korenev}(2011)}]{korenev2011}%
  \BibitemOpen
  \bibfield  {author} {\bibinfo {author} {\bibfnamefont {V.~L.}\ \bibnamefont
  {Korenev}},\ }\href {\doibase 10.1103/PhysRevB.83.235429} {\bibfield
  {journal} {\bibinfo  {journal} {Phys. Rev. B}\ }\textbf {\bibinfo {volume}
  {83}},\ \bibinfo {pages} {235429} (\bibinfo {year} {2011})}\BibitemShut
  {NoStop}%
\bibitem [{\citenamefont {Danon}\ and\ \citenamefont
  {Nazarov}(2008)}]{danon2008}%
  \BibitemOpen
  \bibfield  {author} {\bibinfo {author} {\bibfnamefont {J.}~\bibnamefont
  {Danon}}\ and\ \bibinfo {author} {\bibfnamefont {Y.~V.}\ \bibnamefont
  {Nazarov}},\ }\href {\doibase 10.1103/PhysRevLett.100.056603} {\bibfield
  {journal} {\bibinfo  {journal} {Phys. Rev. Lett.}\ }\textbf {\bibinfo
  {volume} {100}},\ \bibinfo {pages} {056603} (\bibinfo {year}
  {2008})}\BibitemShut {NoStop}%
\bibitem [{\citenamefont {Shi}(2013)}]{shi2013}%
  \BibitemOpen
  \bibfield  {author} {\bibinfo {author} {\bibfnamefont {X.-F.}\ \bibnamefont
  {Shi}},\ }\href {\doibase 10.1103/PhysRevB.87.195318} {\bibfield  {journal}
  {\bibinfo  {journal} {Phys. Rev. B}\ }\textbf {\bibinfo {volume} {87}},\
  \bibinfo {pages} {195318} (\bibinfo {year} {2013})}\BibitemShut {NoStop}%
\bibitem [{\citenamefont {Abragam}(1961)}]{abragam1961}%
  \BibitemOpen
  \bibfield  {author} {\bibinfo {author} {\bibfnamefont {A.}~\bibnamefont
  {Abragam}},\ }\href@noop {} {\emph {\bibinfo {title} {The Principles of
  Nuclear Magnetism}}}\ (\bibinfo  {publisher} {Oxford University Press},\
  \bibinfo {address} {London},\ \bibinfo {year} {1961})\BibitemShut {NoStop}%
\bibitem [{\citenamefont {Yale}\ \emph {et~al.}(2013)\citenamefont {Yale},
  \citenamefont {Buckley}, \citenamefont {Christle}, \citenamefont {Burkard},
  \citenamefont {Heremans}, \citenamefont {Bassett},\ and\ \citenamefont
  {Awschalom}}]{yale2013}%
  \BibitemOpen
  \bibfield  {author} {\bibinfo {author} {\bibfnamefont {C.~G.}\ \bibnamefont
  {Yale}}, \bibinfo {author} {\bibfnamefont {B.~B.}\ \bibnamefont {Buckley}},
  \bibinfo {author} {\bibfnamefont {D.~J.}\ \bibnamefont {Christle}}, \bibinfo
  {author} {\bibfnamefont {G.}~\bibnamefont {Burkard}}, \bibinfo {author}
  {\bibfnamefont {F.~J.}\ \bibnamefont {Heremans}}, \bibinfo {author}
  {\bibfnamefont {L.~C.}\ \bibnamefont {Bassett}}, \ and\ \bibinfo {author}
  {\bibfnamefont {D.~D.}\ \bibnamefont {Awschalom}},\ }\href {\doibase
  10.1073/pnas.1305920110} {\bibfield  {journal} {\bibinfo  {journal}
  {Proceedings of the National Academy of Sciences}\ }\textbf {\bibinfo
  {volume} {110}},\ \bibinfo {pages} {7595} (\bibinfo {year}
  {2013})}\BibitemShut {NoStop}%
\bibitem [{\citenamefont {D'yakonov}\ and\ \citenamefont
  {Perel'}(1974)}]{dyakonov1974}%
  \BibitemOpen
  \bibfield  {author} {\bibinfo {author} {\bibfnamefont {M.}~\bibnamefont
  {D'yakonov}}\ and\ \bibinfo {author} {\bibfnamefont {V.}~\bibnamefont
  {Perel'}},\ }\href@noop {} {\bibfield  {journal} {\bibinfo  {journal} {JETP}\
  }\textbf {\bibinfo {volume} {38}},\ \bibinfo {pages} {177} (\bibinfo {year}
  {1974})}\BibitemShut {NoStop}%
\bibitem [{\citenamefont {Paget}\ \emph {et~al.}(1977)\citenamefont {Paget},
  \citenamefont {Lampel}, \citenamefont {Sapoval},\ and\ \citenamefont
  {Safarov}}]{paget1977}%
  \BibitemOpen
  \bibfield  {author} {\bibinfo {author} {\bibfnamefont {D.}~\bibnamefont
  {Paget}}, \bibinfo {author} {\bibfnamefont {G.}~\bibnamefont {Lampel}},
  \bibinfo {author} {\bibfnamefont {B.}~\bibnamefont {Sapoval}}, \ and\
  \bibinfo {author} {\bibfnamefont {V.~I.}\ \bibnamefont {Safarov}},\ }\href
  {\doibase 10.1103/PhysRevB.15.5780} {\bibfield  {journal} {\bibinfo
  {journal} {Phys. Rev. B}\ }\textbf {\bibinfo {volume} {15}},\ \bibinfo
  {pages} {5780} (\bibinfo {year} {1977})}\BibitemShut {NoStop}%
\bibitem [{\citenamefont {Sladkov}\ \emph {et~al.}(2010)\citenamefont
  {Sladkov}, \citenamefont {Chaubal}, \citenamefont {Bakker}, \citenamefont
  {Onur}, \citenamefont {Reuter}, \citenamefont {Wieck},\ and\ \citenamefont
  {van~der Wal}}]{sladkov2010}%
  \BibitemOpen
  \bibfield  {author} {\bibinfo {author} {\bibfnamefont {M.}~\bibnamefont
  {Sladkov}}, \bibinfo {author} {\bibfnamefont {A.~U.}\ \bibnamefont
  {Chaubal}}, \bibinfo {author} {\bibfnamefont {M.~P.}\ \bibnamefont {Bakker}},
  \bibinfo {author} {\bibfnamefont {A.~R.}\ \bibnamefont {Onur}}, \bibinfo
  {author} {\bibfnamefont {D.}~\bibnamefont {Reuter}}, \bibinfo {author}
  {\bibfnamefont {A.~D.}\ \bibnamefont {Wieck}}, \ and\ \bibinfo {author}
  {\bibfnamefont {C.~H.}\ \bibnamefont {van~der Wal}},\ }\href {\doibase
  10.1103/PhysRevB.82.121308} {\bibfield  {journal} {\bibinfo  {journal} {Phys.
  Rev. B}\ }\textbf {\bibinfo {volume} {82}},\ \bibinfo {pages} {121308}
  (\bibinfo {year} {2010})}\BibitemShut {NoStop}%
\bibitem [{\citenamefont {Danon}(2010)}]{danon2010}%
  \BibitemOpen
  \bibfield  {author} {\bibinfo {author} {\bibfnamefont {J.}~\bibnamefont
  {Danon}},\ }\emph {\bibinfo {title} {Nuclear spin effects in
  nanostructures}},\ \href@noop {} {Ph.D. thesis},\ \bibinfo  {school} {Delft
  University of Technology} (\bibinfo {year} {2010}),\ \bibinfo {note} {the
  derivation of the Fokker Planck equation for the nuclear spin polarization
  can be found in chapter 3. Our Eq.~(\ref{eq:fokkerplanck}) is identical to
  their equation (3.7), but we have expressed $\dot{P}$ in terms of $\delta$,
  $\Gamma_{d}$ and $\Gamma_{h}$.}\BibitemShut {Stop}%
\bibitem [{\citenamefont {Deng}\ and\ \citenamefont {Hu}(2005)}]{deng2005}%
  \BibitemOpen
  \bibfield  {author} {\bibinfo {author} {\bibfnamefont {C.}~\bibnamefont
  {Deng}}\ and\ \bibinfo {author} {\bibfnamefont {X.}~\bibnamefont {Hu}},\
  }\href {\doibase 10.1103/PhysRevB.72.165333} {\bibfield  {journal} {\bibinfo
  {journal} {Phys. Rev. B}\ }\textbf {\bibinfo {volume} {72}},\ \bibinfo
  {pages} {165333} (\bibinfo {year} {2005})}\BibitemShut {NoStop}%
\end{thebibliography}%

\cleardoublepage

\section{Appendix}

\subsection{Lindblad master equation for the driven three-level system}
We present here more extensively our notation and approach for modeling the CPT physics in a driven three-level system. We directly follow Ref.~\cite{fleischhauer2005}. The dynamics of the $\Lambda$~system in Fig.~1(a) is governed by the Hamiltonian (in the rotating frame)
\begin{equation}
H_{\Lambda}=-\frac{\hbar}{2}\begin{pmatrix}
0 & 0 & \Omega_{1}^{*} \\
0 & -8\delta & \Omega_{2}^{*} \\
\Omega_{1} & \Omega_{2} & 2(\Delta-2\delta)
\end{pmatrix}.
\end{equation}
The equation of motion for the density matrix $\rho_{\Lambda}$ that describes this electronic system as an open system with relaxation and decoherence is
\begin{equation}
\dot{\rho}_{\Lambda}=\frac{-i}{\hbar}\left[H_{\Lambda}, \rho_{\Lambda}\right]+\sum_{i,j}\left(L_{ij}\rho_{\Lambda}L_{ij}^{\dagger}-\frac{1}{2}\left\{L_{ij}^{\dagger}L_{ij},\rho_{\Lambda}\right\}\right)\label{eq:eomrho}
\end{equation}
(in our main text, elements $\rho_{ij}$ are density matrix elements of $\rho_{\Lambda}$).
Here, the Lindblad operators are defined by
\begin{subequations}
\begin{equation}
L_{ij}=\alpha_{ij}\ket{i}\bra{j},
\end{equation}
\begin{equation}
\alpha=\frac{1}{2}\begin{pmatrix}
\gamma_{s} & 2\Gamma_{s} & 2\Gamma_{3} \\
2\Gamma_{s} & \gamma_{s} & 2\Gamma_{3}\\
0&0&\gamma_{3}
\end{pmatrix}.
\end{equation}
\end{subequations}
The matrix $\alpha$ contains all decay and decoherence rates of the system: spin flip rate $\Gamma_{s}$, excited state decay rate $\Gamma_{3}$, spin decoherence rate $\gamma_{s}$ and excited state decoherence rate $\gamma_{3}$.
\subsection{Fermi contact hyperfine interaction}
We consider the case where the hyperfine interaction between the $\Lambda$~system and the nuclear spin is dominated by the Fermi contact interaction for the ground state electron. This interaction is described by the Hamiltonian
\begin{equation}
H_{f}=\frac{4}{3}\mu_{0}\mu_{B}\sum_{i}A_{i}\bm{I}_{i}\cdot\bm{S},
\end{equation}
where $A_{i}=\hbar\gamma_{i}|\psi_{e}(\bm{r}_{i})|^{2}$. The gyromagnetic factor, $\gamma_{i}$, and the electron wave function at the position of a nucleus, $\psi_{e}(\bm{r}_{i})$, characterize the interaction strength with the $i$'th nuclear spin. The spin operators are defined to have eigenvalues $m_{J}=-J,\ldots,J$ for any spin quantum number $J$. This interaction term may be viewed in the form of a Zeeman interaction, $H=-\bm{\mu}\cdot\bm{B}_{n}$, with $\bm{\mu}=-g\mu_{B}\bm{S}$ the electron spin magnetic moment. The effective magnetic field due to the nuclei acting on the electron is then
\begin{equation}
\bm{B}_{n}=\frac{4}{3g}\mu_{0}\sum_{i}A_{i}\bm{I}_{i}.
\label{eq:overhauserfield}
\end{equation}
In an external magnetic field it is convenient to expand the $\bm{I}\cdot\bm{S}$ product using ladder operators. The total Hamiltonian becomes
\begin{subequations}
\begin{align}
H&=H_{z}+H_{f},\\
H_{z}&=\hbar\omega_{z}S_{z}+\sum_{i}\hbar\omega_{i}I_{i,z}\label{eq:hz},\\
H_{f}&=\frac{2}{3}\mu_{0}\mu_{B}\sum_{i}A_{i}\left(2I_{i,z}S_{z}+I_{i,+}S_{-}+I_{i,-}S_{+}\right)\label{eq:hf}.
\end{align}
\end{subequations}
Equation~(\ref{eq:hz}) represents the Zeeman energy of the electron spin and the nuclear spins in an external magnetic field applied along $\hat{\mathbf{z}}$. The first term within the summation in Eq.~(\ref{eq:hf}) adds to the external field an effective magnetic (Overhauser) field $B_{n,z}$. To calculate its expectation value $\braket{B_{n,z}}=\mathrm{Tr}(B_{n,z}\rho_{n})$, where $\rho_{n}$ is the reduced density matrix comprising the nuclear spin state, it is in principle required to know the interaction strengths for all nuclei. In the case of GaAs this is well studied and $\braket{B_{n,z}}\approx  \braket{{I}_{z}} \cdot 3.53~{\rm T}$ \cite{paget1977}, and the maximum field is $B_{\text{max}}=5.30~{\rm T}$. The Overhauser field $B_{n,z}$ translates to the Overhauser shift $\delta$ used in the main text according to
$\delta = \frac12 g\mu_{B} B_{n,z}/\hbar$. This yields $\delta_{\text{max}}=16.3$~GHz.
To describe DNP we use a so-called box model \cite{urbaszek2013} where the eletron couples equally to a number of $N$ nuclear spins. This amounts to the change $\sum_{i}A_{i}\rightarrow A\sum_{i=1}^{N}$ with $A$ the average interaction strength per nucleus. In our calculations we approximate GaAs by choosing $N=10^5$.

The constant  $K$ in Eq.~(2) is
\begin{equation}
K=\frac{4\mu_{0}\mu_{B}}{3\hbar}\sum_{i}A_{i}\frac{I_{i}^{2}+I_{i}}{S^{2}+S}.
\label{eq:constantK}
\end{equation}
For GaAs, $I_{i}=3/2$ for all nuclei. So $K=10\delta_{\text{max}}/3=54.3$~GHz.
\subsection{Hyperfine relaxation rate}
The cross relaxation between the electron spin and the nuclear spins is facilitated by a modulation of the hyperfine coupling due to random jumps in the electron spin state. These jumps occur on average after a correlation time $\tau_{c}$. The relaxation rate is then the product of the average hyperfine coupling, the fraction of time the electron is present ($f_{e}$) and the spectral density of the electron spin fluctuations \cite{abragam1961,urbaszek2013},
\begin{equation}
\Gamma_{h}=\left(\frac{A}{N\hbar}\right)^2 2f_{e} \frac{\tau_{c}}{1+(\omega_{z}+\delta)^2 \tau_{c}^2}.
\end{equation}
The relaxation process of the $\hat{\mathbf{z}}$ projection of the nuclear spin is allowed due to jumps in the perpendicular component of $\mathbf{S}$. For the undriven electron spin $\tau_{c}$ equals $T_{2}=1/\gamma_{s}$, \textit{i.e.} the intrinsic decoherence time of the electron spin. Under conditions of laser driving $\tau_{c}$ is reduced when the laser driving leads to repeated excitation and spontaneous emission. The sharp variation of absorption around CPT has to be taken into account in our model. To deal with this we assume that we operate under conditions where $\omega_{z} \gg \delta$ and $\omega_{z} \gg 1/\tau_{c}$ so that the spectral density is approximately proportional to the inverse correlation time
\begin{equation}
\frac{\tau_{c}}{1+(\omega_{z}+\delta)^2 \tau_{c}^2}\approx\frac{1}{\omega_{z}^2 \; \tau_{c}}.
\end{equation}
In addition, we take the inverse correlation time to be enhanced by the amount of optical transitions that disturb the electron spin state, which we can obtain from the $\Lambda$ system model, \textit{i.e.} $1/\tau_{c}=(\rho_{11}+\rho_{22})\gamma_{s}+\rho_{33}\Gamma_{e}$.

\begin{figure}
\centering
\includegraphics[width=\columnwidth]{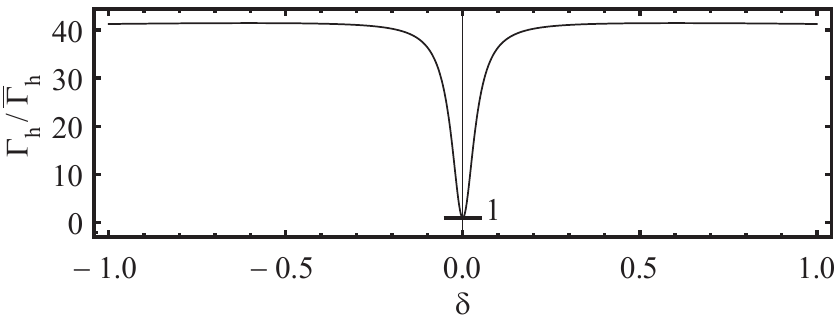}
\caption{Modulation of the hyperfine relaxation rate $\Gamma_{h}$ by the Overhauser shift $\delta$ under conditions of two-laser driving. Detuning $\Delta=1$, parameters $\Omega_{1,2}$, $\Gamma_{s}$, $\gamma_{s}$, $\Gamma_{3}$, $\gamma_{3}$ are as in Fig.~1. This graph has been used for the calculations for Figs.~2--4.}
\label{fig:fliprate}
\end{figure}

In our simulations we specify a value for $\overline{\Gamma}_{h}/\Gamma_{d}$ (this value is reported in the captions of Figs.~2--4) where $\overline{\Gamma}_{h}$ is the hyperfine relaxation rate of the equilibrium system (no laser driving). 
This provides the basis for the effective value of $\Gamma_{h}$, for which we can calculate its dependence on $\delta$ through $\tau_{c}$.
How this dependence controls a modulation of the effective value for $\Gamma_{h}/\overline{\Gamma}_{h}$ near CPT conditions is presented in Fig.~\ref{fig:fliprate} for a specific set of optical driving parameters (see caption).
\subsection{Steady state solution to the Fokker-Planck equation}
A steady state ($\dot{P}=0$) solution to Eq. 3 is 
\begin{equation*}
P_{ss}(\delta)=\eta~\mathrm{exp}\left(-\int_{0}^{\delta}f_{1}(x)/f_{2}(x)\mathrm{d}x\right),
\end{equation*}
where
\begin{align*}
f_{1}(x)&=-\dot{\delta}(x)+\delta_{\text{max}}^{2}\frac{\partial}{\partial x}(\Gamma_{d}+\Gamma_{h}(x))/N,\\
f_{2}(x)&=\delta_{\text{max}}^{2}(\Gamma_{d}+\Gamma_{h}(x))/N
\end{align*}
and $\eta$ is a number that is fixed by the normalization condition $\int P(\delta)~\mathrm{d}\delta=1$. A special solution arises in the case when $f_{1}(x)=a x$ and $f_{2}(x)=b$ with $a,b$ constant. Then the steady state distribution is Gaussian with standard deviation $\sigma=(b/a)^{1/2}$. 
\subsection{Electron-spin dephasing from hyperfine interaction with a nuclear spin bath}
Because of the slow dynamics of the nuclear spins compared tot the electron spin, each measurement on the electron spin is subject to an Overhauser field $B_{n,z}$ sampled from a distribution. For example, at thermal equilibrium at the high temperatures that we consider (for nuclear spins), this is a Gaussian distribution with mean $\braket{B_{n,z}}=0$ and standard deviation $\sigma_{B}$. For a measurement on an ensemble of electron spins (or many separate single spin measurements), one will observe inhomogeneous dephasing as a function of time $t$. This can be parameterized with a function $C(t)$ that evolves from no dephasing to complete dephasing on a scale from 1 to 0:
\begin{equation}
C(t)=\left|\int_{-\infty}^{+\infty}P(B) \; \exp\left(-\frac{ig\mu_{B}Bt}{\hbar}\right)\mathrm{d}B \right|.
\label{eq:dephasing}
\end{equation}
Here $P(B)$ is the probability distribution for the total field $B = B_{ext}+B_{n,z}$ (where $B_{ext}$ is the externally applied magnetic field), taken over an ensemble of electrons. This expression captures the gradual loss of information about $S_{x}$ and $S_{y}$ as a function of time. For the Gaussian distribution at thermal equilibrium
\begin{equation}
\overline{P}(B)=\frac{1}{\sqrt{2\pi\sigma_{B}^2}} \; \exp\left(-\frac{B^2}{2\sigma_{B}^{2}}\right).
\end{equation}
The dephasing time scale $T_{2}^{*}$ is defined as the time where Eq.~(\ref{eq:dephasing}) reduces to $1/\mathrm{e}$.
For the Gaussian distribution $\overline{P}(B)$, Eq.~(\ref{eq:dephasing}) yields $C(t)$ in the form $\exp\left[-(t/T_{2}^{*})^{2}\right]$ with the inhomogeneous dephasing time
\begin{equation}
T_{2}^{*}=\frac{\sqrt{2}\hbar}{|g|\mu_{B}\sigma_{B}}.
\end{equation}
The steady state distributions obtained from the feedback model with nonlinear response are not Gaussian, for those no simple expression for $T_{2}^{*}$ is available. We define $T_{2}^{*}$ as the time at which $C(t)$ has dropped to $1/\mathrm{e}$ of its initial value, which is obtained by numerical evaluation of Eq. \ref{eq:dephasing}.
Further, it is straightforward to calculate with this definition a value for $T_{2}^{*}$ for any of the distributions $P(\delta)$ that is presented in the main text (using $\delta = \frac12 g\mu_{B}B_{n,z}/\hbar$).

\end{document}